\documentclass[runningheads, envcountsect]{llncs}
\usepackage[utf8]{inputenc}
\usepackage[T1]{fontenc}
\usepackage{graphicx}
\usepackage{mathtools}

\usepackage{amsmath}
\usepackage{amssymb}

\spnewtheorem{thm}[theorem]{Theorem}{\bfseries}{\itshape}
\spnewtheorem{rmk}[theorem]{Remark}{\bfseries}{\upshape}
\spnewtheorem{defn}[theorem]{Definition}{\bfseries}{\itshape}
\spnewtheorem{exmp}[theorem]{Example}{\itshape}{\upshape}

\usepackage[hidelinks, linktoc=all]{hyperref}

\usepackage{color}

\usepackage{textcomp}
\usepackage{textgreek}
\usepackage{upgreek}
\usepackage{mathrsfs}
\usepackage{xspace}

\usepackage{listings}

\usepackage{xcolor}
\definecolor{keywordcolor}{rgb}{0.7, 0.1, 0.1}   
\definecolor{commentcolor}{rgb}{0.4, 0.4, 0.4}   
\definecolor{symbolcolor}{rgb}{0.0, 0.1, 0.6}    
\definecolor{sortcolor}{rgb}{0.1, 0.5, 0.1}      

\providecommand{\href}[2]{#2}

\providecommand*{\backref}{}
\providecommand*{\backrefalt}{}
\renewcommand*{\backref}[1]{}
\renewcommand*{\backrefalt}[4]{%
	\ifcase #1 %
	\or
	  Cited page~#2.
	\else
	  Cited pages~#2.
	\fi
}

\newcommand\MTkillspecial[1]{
  \bgroup
  \catcode`\&=9
  \let\\\relax%
  \scantokens{#1}%
  \egroup
}
\newcommand\DeclarePairedDelimiterMultiline[3]{
  \DeclarePairedDelimiter{#1}{#2}{#3}
  \reDeclarePairedDelimiterInnerWrapper{#1}{star}{
    \mathopen{##1\vphantom{\MTkillspecial{##2}}\kern-\nulldelimiterspace\right.}
    ##2
    \mathclose{\left.\kern-\nulldelimiterspace\vphantom{\MTkillspecial{##2}}##3}}
}

\newcommand{\Z}{\mathbb{Z}}
\newcommand{\N}{\mathbb{N}}
\newcommand{\R}{\mathbb{R}}

\newcommand{\dd}{\mathop{}\!\mathrm{d}}
\DeclarePairedDelimiterMultiline{\abs}{\lvert}{\rvert}
\DeclarePairedDelimiterMultiline{\norm}{\lVert}{\rVert}

\DeclareMathOperator{\Leb}{Leb}
\DeclareMathOperator{\dLeb}{dLeb}
\newcommand{\coloneqq}{\mathrel{\mathop:}=}

\renewcommand{\epsilon}{\varepsilon}

\renewcommand{\phi}{\varphi}
\renewcommand{\leq}{\leqslant}
\renewcommand{\geq}{\geqslant}
\newcommand{\mathlib}{\texttt{mathlib}\xspace}

\begin{document}
\title{A formalization of the change of variables formula for integrals in mathlib}
\titlerunning{Change of variables in mathlib}
%
\author{Sébastien Gouëzel\inst{1}\orcidID{0000-0002-7188-8671}}
\authorrunning{S. Gouëzel}
%
\institute{IRMAR, CNRS UMR 6625, Université de Rennes 1, 35042 Rennes, France
\email{sebastien.gouezel@univ-rennes1.fr}}
\maketitle              
\begin{abstract}
We report on a formalization of the change of variables formula in
integrals, in the mathlib library for Lean. Our version of this theorem is
extremely general, and builds on developments in linear algebra, analysis,
measure theory and descriptive set theory. The interplay between these
domains is transparent thanks to the highly integrated development model of
mathlib.

\end{abstract}
\section{Introduction}

The change of variables formula in integrals is a basic tool in mathematics,
playing an important role both in concrete computations of integrals and in
more theoretical domains, notably Poincaré duality for de Rham cohomology.
Its most basic formulation is the following:

\begin{thm}
\label{thm:main} Consider the vector space $\R^n$ with it standard Lebesgue
measure, and $f : \R^n \to \R^n$ a $C^1$-diffeomorphism (i.e., $f$ is a
bijection, it is continuously differentiable, and so is its inverse). Then,
for any integrable function $g : \R^n \to \R$, the function $x \mapsto
\abs{\det Df(x)} \cdot g(f (x))$ is also integrable and
\begin{equation*}
  \int g(y) \dLeb(y) = \int \abs{\det Df(x)} \cdot g(f (x)) \dLeb(x).
\end{equation*}
\end{thm}

This paper is devoted to the description of a formalization of (a more
sophisticated version of) this theorem, in the Lean proof assistant,
developed at Microsoft Research by Leonardo de Moura~\cite{demoura_lean},
within the library \mathlib~\cite{mathlib}. Apart from its mathematical
relevance, an interest of this theorem from the formalization point of view
is that it mixes several domains of mathematics that are typically taught in
different courses, notably linear algebra, calculus, measure theory (and
descriptive set theory for the aforementioned more sophisticated version).
Therefore, it can only be formalized in a library which is developed enough
in all these directions, and in which all these areas can interact in a
coherent way. This is the case of \mathlib, but also of the main library of
Isabelle/HOL (which already contains a version of the above theorem) or of
\texttt{mathcomp-analysis} in Coq (which does not contain a version of the
above theorem at the time of this writing, but might in the near future). The
need for such coherence in advanced mathematics libraries will be a guiding
theme in this paper.

This paper is written both for mathematicians who want to learn more on
advanced versions of the change of variables formula or on theorem provers,
and for formalizers: we will explain design issues that show up at different
places, and justify the specific choices that have been made in the \mathlib
formalization to solve these issues.

\section{Sketch of proof of Theorem~\ref{thm:main}}
\label{sec:sketch}

Let us sketch a proof of Theorem~\ref{thm:main} as may be found in standard
textbooks, to highlight the tools that are needed. Approximating the function
$g$ by characteristic functions of measurable sets, and then the measurable
set by a compact set, it is sufficient to prove the following statement: if
$K$ is a compact set, then
\begin{equation*}
  \Leb(f(K)) = \int_K \abs{\det Df(x)} \dLeb(x).
\end{equation*}
We will check that each of these quantities is bounded above by the other
one.

Fix $\delta>0$. Let $\epsilon>0$. Cover $K$ by boxes made from a grid of mesh
$\epsilon$, and denote the center of such a box $B_i$ by $c_i$. By uniform
continuity of $Df$ on compact sets, if $\epsilon$ is small enough, the differential $Df$ is
arbitrarily close to $Df(c_i)$ on $B_i$. It follows that $f(B_i)$ is included
in the image of $B_i$ under the linear map $(1+\delta)Df(c_i)$, uniformly in
$i$. Then
\begin{align*}
  \Leb(f(K)) &\leq \sum_i \Leb(f(B_i)) \leq \sum_i \Leb((1+\delta)Df(c_i)(B_i))
  \\&
  = (1+\delta)^n \sum_i \abs{\det (Df(c_i))} \Leb(B_i),
\end{align*}
where the last equality follows from the fact that a matrix $A$ rescales the
volume according to $\abs{\det A}$. If $\delta$ is small enough, then $
\abs{\det (Df(c_i))} \leq (1+\delta) \abs{\det Df(x)}$ for any $x\in B_i$ by
uniform continuity. Then the above sum can be bounded by
\begin{equation*}
  (1+\delta)^n \sum_i \int_{B_i} (1+\delta) \abs{\det Df(x)} \dLeb(x).
\end{equation*}
Finally, denoting by $K_\epsilon$ the neighborhood of $K$ given by the union
of the $B_i$, we have proved that
\begin{equation*}
  \Leb(f(K)) \leq (1+\delta)^{n+1} \int_{K_\epsilon} \abs{\det Df(x)} \dLeb(x).
\end{equation*}
When $\epsilon$ tends to $0$, then $K_\epsilon$ tends to $K$. The dominated
convergence theorem shows that the integral over $K_\epsilon$ converges to
the integral over $K$. Finally, as $\delta$ is arbitrary, we have proved the
inequality
\begin{equation*}
  \Leb(f(K)) \leq \int_K \abs{\det Df(x)} \dLeb(x).
\end{equation*}

The converse inequality can be proved along the same lines, but one step is
more delicate: one should show that $\Leb(f(B_i)) \geq (1+\delta)^{-n}
\Leb(Df(c_i)(B_i))$, i.e., one should show that $f(B_i)$ is comparatively
large. A new ingredient is needed there, to prove that $f$ is locally
surjective (in a quantitative way). This follows from the inverse function
theorem. Equivalently, one can mimic the above computation but for the map
$f^{-1}$ (which is also a diffeomorphism). This concludes the proof. \qed

\section{A more sophisticated version of the theorem}

From the proof sketch above, it is obvious that some assumptions in
Theorem~\ref{thm:main} may be relaxed. For instance, it is not necessary that
$f$ is defined on the whole space: if it is a diffeomorphism between two open
sets of $\R^n$, then the same proof will go through. It is also not necessary
that the vector space is $\R^n$: any finite-dimensional real vector space
with a Lebesgue measure (i.e., a translation invariant sigma-finite nonzero measure)
will do, as such a measure has the same rescaling properties under linear
maps.

On the other hand, since the proof relies on the inverse function theorem, it
looks as though assuming that the map is defined on an open set and its
differential is continuous and invertible can not be avoided. It turns out
that this is not the case (and this may come as a surprise even to
mathematicians who are very familiar with this theorem and its applications):
all these assumptions can be dispensed with.

The most general version of the theorem is expressed in terms of a (slightly
non-standard) notion of differentiability along a set:
\begin{defn}
\label{defn:differentiable} Consider a normed real vector space $E$, a map $f
: E \to E$ and a continuous linear map $A : E \to E$. We say that $A$ is a
derivative of $f$ at a point $x \in E$ along a set $s \subseteq E$ if, when
$y$ tends to $x$ inside $s$, then $f(y) = f(x) + A(y-x) + o(y-x)$.
\end{defn}
When $s$ is the whole space (or a neighborhood of $x$), this coincides with
the usual notion of differentiability at $x$.

We say that a measure on a finite-dimensional real vector space is a
\emph{Lebesgue measure} if it is nonzero, sigma-finite, and invariant under
left-translation. Such a measure always exists, and it is unique up to scalar
multiplication. Such measures are also known as Haar measures in the more
general context of locally compact topological groups.

Here is the general version of the change of variables formula.
\begin{thm}
\label{thm:main2} Let $E$ be a finite-dimensional normed real vector space
endowed with a Lebesgue measure, $f : E \to E$ a map, $s$ a Borel-measurable
subset of $E$ and $f'(x)$ a linear map on $E$ for each $x$. Assume that $f$
is injective on $s$, and that at every $x \in s$ the linear map $f'(x)$ is a
derivative of $f$ at $x$ along $s$. Then $f(s)$ is also Borel-measurable, and
any function $g : E \to \R$ satisfies the equality
\begin{equation*}
  \int_{f(s)} g(y) \dLeb(y) = \int_s \abs{\det f'(x)} \cdot g(f(x)) \dLeb(x).
\end{equation*}
\end{thm}
As in the proof sketch of Paragraph~\ref{sec:sketch}, this follows from a
result on the measure of the image set:
\begin{equation}
\label{eq:Lebfs_hard}
  \Leb(f(s)) = \int_s \abs{\det f'(x)} \dLeb(x).
\end{equation}

This theorem is proved in~\cite[Theorem 263D]{fremlin2}, with the difference
that the emphasis there is on Lebesgue-measurable sets more than
Borel-measurable ones. To obtain the fact that $f(s)$ is Borel-measurable if
$s$ is, one needs additionally the Lusin-Souslin theorem~\cite[Theorem
423I]{fremlin4}, an important and nontrivial result in descriptive set
theory.

We will not give a full proof of Theorem~\ref{thm:main2}, and refer the
interested reader to~\cite{fremlin2} instead. Let us only stress where one
can follow the sketch given in Section~\ref{sec:sketch}, and where one should
depart from it. Let us focus on the proof of~\eqref{eq:Lebfs_hard}. Fix a
small $\epsilon>0$. With the definition of differentiability along $s$, one
may split $s$ into countably many disjoint small sets $s_i$ on which $f$ is
well approximated by a linear map $A_i$, up to $\epsilon$. Then one would
like to say that $\Leb(f(s_i))$ is close to $\abs{\det (A_i)} \cdot
\Leb(s_i)$, but one can not resort to the inverse function theorem.

For the direct inequality
\begin{equation}
  \label{eq:Lebfsile}
  \Leb(f(s_i)) \le (\abs{\det (A_i)} + \epsilon) \cdot \Leb(s_i),
\end{equation}
we fix $\delta>0$ and we use a covering lemma (such as the Vitali or the
Besicovitch covering theorems -- see Section~\ref{sec:covering} for
more on these theorems) ensuring that one can cover $s_i$ with
countably many balls $(B_{ij})_{j\in \N}$ whose measures add up to at most
$\Leb(s_i) + \delta$. For each such ball $B_{ij}$, its image has measure
bounded by $(\abs{\det (A_i)} + \epsilon) \Leb(B_{ij})$. Adding these
estimates and letting $\delta$ tend to $0$, we get~\eqref{eq:Lebfsile}.

For the converse inequality
\begin{equation*}
  (\abs{\det (A_i)} - \epsilon) \cdot \Leb(s_i) \le \Leb(f(s_i)),
\end{equation*}
there is nothing to prove if $A_i$ is not invertible. If it is invertible,
one argues that $f$ is invertible on $s_i$ and that its inverse is close to
$A_i^{-1}$ (this is a nonstandard version of the inverse function theorem).
Then, one repeats the above computation for $f^{-1}$ to get the desired
estimate.

Adding all these inequalities over $i$ one gets that $\Leb(f(s))=\sum_i
\Leb(f(s_i))$ is comparable to $\sum \abs{\det (A_i)} \Leb(s_i)$, which is
comparable to $\int_s \abs{\det f'(x)} \dLeb(x)$ as $f'(x)$ is close to $A_i$
on $s_i$. This concludes the proof. \qed

\medskip

A difficulty that we have ignored in this proof sketch is that the derivative
along a set is in general not unique at points where the set is not fat
enough. This has the unpleasant consequence that, in the statement of
Theorem~\ref{thm:main2}, the function $\abs{\det f'(x)}$ is in general not
Borel-measurable along $s$, as illustrated by the following example.

\begin{exmp}
Take $s = \R \times \{0\} \subseteq \R^2$, and $f(x,y) = (x, 0)$. Let also
$t$ be any (possibly non-measurable) subset of $\R$, and set
\begin{equation*}
  f'(x, y) = \left(\begin{matrix} 1 & 0 \\ 0 & 1_t(x)\end{matrix}\right).
\end{equation*}
Along horizontal directions, $f'(x, y)$ acts as the identity, so it is indeed a
derivative of $f$ along $s$. But $\det f'(x,y) = 1_t(x)$ is not a measurable
function.
\end{exmp}
Nevertheless,  Theorem~\ref{thm:main2} is still true in this example, since
$s$ and $f(s)$ have zero measure, so both the left and the right hand side in the statement
of the theorem
vanish. In general, $f'(x)$ is uniquely defined on a full measure subset of
$s$ (its \emph{Lebesgue density points}, which are again studied using
covering lemmas), and is measurable there. In particular, even though
$\abs{\det f'(x)}$ is not always Borel-measurable, it coincides almost
everywhere with a Borel-measurable function.

For Theorem~\ref{thm:main2} to be true, it means that the theory of
integration one uses should work smoothly with functions which are not
Borel-measurable, but coincide almost everywhere with a Borel-measurable
function. While this was not the case of the first definition of the integral
in \mathlib, it had already been refactored (for different reasons) before
the start of this project to allow non-Borel measurable functions, so no
modification of the library was needed on this side.

Let us explain why the definition of integral had been refactored prior to
this work. \mathlib initially contained a definition of the integral for
which the integral of non-measurable functions was zero by convention. This
seemed quite satisfactory and made it possible to prove many theorems, until
the formalization of the fundamental theorem of calculus. This theorem reads
as follows: if two functions $f, g : \R \to \R$ are continuous on an interval
$[a, b]$ and $g$ is the derivative of $f$ there, then $\int g(x)
\dLeb_{|[a,b]}(x) = f(b) - f(a)$. It turns out that this theorem as stated
was not true in \mathlib: we are not making any assumption on $g$ outside
$[a,b]$, so there is no reason why $g$ should be measurable globally, which
means that $\int g(x) \dLeb_{|[a,b]}(x)$ could be equal to $0$ for no good
reason. The first version of this theorem in \mathlib therefore needed an
additional assumption that $g$ was measurable globally.

Without this assumption, it is true that $g$ is null-measurable (i.e.\ almost
everywhere measurable) with respect to the restricted measure
$\dLeb_{|[a,b]}$: it coincides almost everywhere with a measurable function,
namely the function equal to $g$ on $[a,b]$ and to $0$ elsewhere (which is
indeed measurable by continuity of $g$ on $[a,b]$).

To get a more satisfactory statement for the fundamental theorem of calculus,
the definition of the integral in \mathlib was therefore refactored to allow
for almost everywhere measurable functions: if a function is almost
everywhere measurable, then its integral is defined to be the integral of a
measurable function which coincides with it almost everywhere, and otherwise
the integral is defined to be $0$.

This definition has several advantages. For instance, if two functions
coincide almost everywhere then they have the same integral regardless
of measurability issues. Moreover, it is
exactly the kind of integration theory which is needed for
Theorem~\ref{thm:main2} to hold! That this change was needed and fruitful would not have
been noticed in a pure measure-theory library, and was really a consequence
of the interaction of different domains of mathematics in \mathlib.

\emph{Getting definitions right the first time is hard. Definitions should be
driven by the theorems they enable, even in different domains, and one should
not be afraid to refactor a core definition.}

As an aside, let us note that the definition of integrals was refactored a
third time in \mathlib, to allow for functions that take values in spaces
which are not second-countable. While the standard definition of integration
(writing a function as a pointwise limit of simple functions) is easier to
work out when the target space is second-countable, this restriction prevents
some applications to complex analysis and spectral theory. When these
limitations were noticed, the definition was changed again, for the better.
Now, the functions that can be integrated in \mathlib are the almost
everywhere strongly measurable ones, i.e., the functions that coincide almost
everywhere with a pointwise limit of simple functions. And there are several
results ensuring that most concrete functions are almost everywhere strongly
measurable functions -- for instance measurable functions into
second-countable spaces, or continuous functions from second-countable
spaces.

\section{The formalized version of the theorem}

Here is the full statement of the formalized version of
Theorem~\ref{thm:main2}.

\begin{lstlisting}
theorem integral_image_eq_integral_abs_det_fderiv_smul
[normed_group E] [normed_space ℝ E] [finite_dimensional ℝ E]
[measurable_space E] [borel_space E]
(μ : measure E) [is_add_haar_measure μ]
[normed_group F] [normed_space ℝ F] [complete_space F]
{s : set E} {f : E → E} {f' : E → (E →L[ℝ] E)} (hs : measurable_set s)
(hf' : ∀ x ∈ s, has_fderiv_within_at f (f' x) s x)
(hf : set.inj_on f s) (g : E → F) :
∫ y in f '' s, g y ∂μ = ∫ x in s, |(f' x).det| • g (f x) ∂μ
\end{lstlisting}

Here is a rephrasing of the theorem for readers who are not familiar with
Lean's syntax. We start with a finite-dimensional real normed vector space
$E$, a measure $\mu$ on $E$ which is assumed to be a Lebesgue measure (in
more formal terms, an additive Haar measure), a subset $s$ of $E$ which is
assumed to be measurable (assumption \verb+hs+), and a function $f : E \to E$
which is injective on $s$ (assumption \verb+hf+). Consider also, for each $x
\in E$, a continuous linear map $f'(x)$ on $E$. Assume that, for each $x \in
s$, then $f'(x)$ is a derivative of $f$ at $x$ along $s$ (assumption
\verb+hf'+). Then the change of variables formula holds: for any function $g
: E \to F$ (where $F$ is any complete real vector space), then
\begin{equation*}
  \int_{y\in f(s)} g(y) \dd\mu(y) = \int_{x\in s} \abs{\det(f'(x))} g(x) \dd\mu(x).
\end{equation*}
This corresponds perfectly to Theorem~\ref{thm:main2}.

Let us list the different domains of mathematics that are involved in the
statement of Theorem~\ref{thm:main2}:
\begin{enumerate}
\item Analysis and topology: to talk about normed spaces and continuity.
\item Calculus: to make sense of derivatives.
\item Measure theory: to talk about integrals and measures of sets
    (including the definition of additive Haar measures).
\item Linear algebra: to talk about finite dimensional spaces, and also
    about determinants of linear maps.
\end{enumerate}
All these domains should be formalized before the above formalized statement
\verb+integral_image_eq_integral_abs_det_fderiv_smul+ can be merely written
down and understood by the system.

There are also tools that show up in the proof of the theorem, but not in its
statement:
\begin{enumerate}\setcounter{enumi}{4}
\item Ordinals and transfinite induction (these show up in the proof of the
    covering theorems).
\item Linear maps rescale Lebesgue measures according to the absolute value
    of their determinants.
\item Covering theorems, like the Besicovitch and Vitali covering theorems.
\item Descriptive set theory, notably the theory of Polish spaces and
    analytic sets in them (they are instrumental in the proof of the
    Lusin-Souslin theorem).
\end{enumerate}
All these should be formalized before the proof of the change of variables
theorem.

This project resulted in 80 pull requests to \mathlib, adding roughly 15,000
lines of code. Among these, most are devoted to the prerequisites presented
above: the file on the change of variables formula itself has only 1259
lines, less than 10\% of the total. In the topics above, Items 1--5 were
already mature enough that they needed few additions. Items 6--8 form the
bulk of the formalization of this project, with roughly 20\% for 6, the
remaining 70\% being split evenly between 7 and 8.

\mathlib is an open source project: everyone can submit pull requests, which
are then submitted to a thorough review process. There are 25 maintainers
of the project. When a maintainer is happy with a pull request (and several
sanity checks have been automatically performed, as explained
in~\cite{mathlib_lint}), then he can merge it to the main branch (and of
course no maintainer can merge his own pull requests). Given the width of
\mathlib, no maintainer is expert in all areas: the pull requests in this
project were therefore refereed by different maintainers depending on their
domains. An important point is that the maintainers coordinate to ensure the
unity of the whole library. For instance, the linear maps that are used in
linear algebra are the same as those that are used in algebraic applications
such as Galois theory, or in analytic applications such as derivatives of
maps.

This inter-operability is extremely useful for a project such as the change
of variables formula, that involves many different areas of mathematics: in a
less coherent project, one would likely need to add glue to make sure that
different modules can work together, and this would become quickly unwieldy
at this level of complexity. In this respect, in the language
of~\cite{cathedral_bazaar}, the \mathlib library is cathedral-like as it is
complex and coherent, but its open-source development process also has some
bazaar characteristics. The delicate balance between these two models is only
possible thanks to the hard work of the maintainers, who should be thanked
for their dedication.

The next three sections will be devoted to more in-depth discussion of the
three main ingredients 6--8. Before that, let us make a few remarks on the
formalized statement of the theorem.

\begin{rmk}
There are several assumptions in this theorem that appear between brackets,
like \verb+[normed_group E]+. These are \emph{typeclass assumptions}, that
should be filled automatically by the system when the theorem is used. The
only assumptions that should be checked by the user are those between
parentheses, like \verb+(hs : measurable_set s)+. The typeclass assumptions
are checked by the system using special lemmas that are tagged as
\emph{instances}. For instance, the fact that the volume on $\R$ is a
Lebesgue measure is an instance, as well as the fact that the product of two
Lebesgue measures is a Lebesgue measure, so the theorem will automatically
apply to the standard Lebesgue measure on $E = \R \times \R$ (and the fact
that $\R \times \R$ is a finite-dimensional real normed vector space is also
checked automatically by typeclass inference).
\end{rmk}

\begin{rmk}
In this version of the theorem, $E$ is a general finite-dimensional real
vector space, and $\mu$ is a general Lebesgue measure on $E$ (this is the
content of the typeclass assumption \verb+[is_add_haar_measure μ]+). One may
wonder if one really needs this generality, and if it would not be more
natural to have the theorem only on $\R^n$ with its canonical (product)
Lebesgue measure. For instance, this is the way the analogue of this theorem
is formulated in Isabelle/HOL. However, we found that this more restricted
version is too limited.

As an illustration, let us recall a classical proof of the value of the
Gaussian integral $\int_\R e^{-x^2/2} \dd x = \sqrt{2\pi}$. Denoting by $I$
the value of the integral, one can compute using a polar change of
coordinates in $\R \times \R$, writing $(x,y)=(r\cos \theta, r \sin \theta)$
as follows:
\begin{align*}
  I^2 & = \int_{\R \times \R} e^{-(x^2 + y^2)/2} \dd x \dd y
  = \int_{r=0}^{+\infty} \int_{\theta = -\pi}^\pi e^{-r^2/2} \cdot r \dd r \dd \theta
  \\& = 2\pi \int_{r=0}^\infty r e^{-r^2/2} \dd r
  = 2\pi,
\end{align*}
where the computation $\int_{r=0}^\infty r e^{-r^2/2} \dd r = 1$ follows from
the fact that $r e^{-r^2/2}$ is the derivative of $e^{-r^2/2}$.

This proof (which has been formalized in \mathlib) uses the change of
variables theorem in the space $E=\R \times \R$, with the product Lebesgue
measure. But this space is \emph{not} one of the spaces $\R^n$, defined as
the space of functions from $\{0, \dotsc, n-1\}$ to $\R$: the product space
$\R \times \R$ and the function space $\{0,1\} \to \R$ are obviously the same
for a mathematician, but in all rigor they are different (albeit canonically
isomorphic), and in particular they are definitely different from Lean's
point of view. One could try to reformulate the above proof using $\R^2$
instead of $\R \times \R$, but then Fubini theorem (which has been used in
the first step of the above proof) would not apply directly as it only makes
sense for product spaces.

This is an important lesson: \emph{more general theorems apply in more
situations, so one should aim for generality in formalized mathematics
library to improve their usability.}
\end{rmk}

\begin{rmk}
The differentiability assumption inside a set is not completely standard in
mathematics, but it shows up often in particular cases. Notably, in one
dimension, one often talks about left derivatives and right derivatives,
which are particular cases of Definition~\ref{defn:differentiable} where $s$
is respectively $(-\infty, x]$ or $[x, +\infty)$. In differential geometry,
one also often encounters functions which are differentiable on half-spaces
or on submanifolds. With these examples in mind, derivatives in \mathlib were
defined from the start using Definition~\ref{defn:differentiable}, so they
were already general enough for Theorem~\ref{thm:main2}. As most of the
basics of analysis and topology in \mathlib, this is strongly inspired
from the Isabelle/HOL formalization of analysis described
in~\cite{analysis_HOL}.
\end{rmk}

\begin{rmk}
There is a difference between the statement of Theorem~\ref{thm:main2} and
its formalized version: in Theorem~\ref{thm:main2}, we require the function
$g$ to be integrable on $f(s)$, but this assumption is nowhere to be seen in
the formalized version (not even measurability or almost everywhere
measurability of $g$). This makes the formalized version easier to use: the
user does not need to worry about proving the integrability of the function.

The formalized version comes with a companion theorem, saying that $g$ is
integrable on $f(s)$ if and only if $x\mapsto \abs{\det f'(x)} \cdot g(x)$ is
integrable on $s$. Since the
integral of non-integrable functions is defined to be $0$ by convention, the
theorem is then tautologically true for non-integrable functions as both
sides of the statement vanish.

The companion theorem is not trivial, especially regarding measurability
issues. Ultimately, it relies on the fact that the restriction of $f$ to $s$
is a measurable embedding, i.e., it maps measurable sets to measurable sets.
This follows from the deep Lusin-Souslin theorem in descriptive set theory. A
first version of the formalization had the additional assumption that $g$ was
integrable and avoided the Lusin-Souslin theorem. It was less satisfactory
since it required more work from the end user of the theorem when applying
it.

This rule is followed throughout \mathlib: \emph{One should try to minimize
the assumptions of theorems to make them easier to apply for the users, even
if this comes with a higher proof burden for the formalizer of the theorem}.
(In our specific case, the Lusin-Souslin theorem indeed required several
thousands additional lines of formalization.)

Let us give another silly example of this rule: the formula $\int f + g =
\int f + \int g$ requires the functions $f$ and $g$ to be integrable. On the
other hand, the formula $\int cf = c \int f$ is true whether $f$ is
integrable or not, so the latter formula should be given without
integrability assumptions, even though the proof becomes more complicated as
it requires several case distinctions. Indeed, if $f$ is nonintegrable and
$c$ is nonzero, then $cf$ is also nonintegrable so both sides vanish; if $f$
is nonintegrable and $c$ is zero, then $cf$ becomes integrable as it is the
zero function, but both sides vanish again; if $f$ is integrable then $cf$ is
also integrable and one is back to the usual situation.
\end{rmk}

\section{Linear maps rescale Lebesgue measure according to their determinants}

A basic ingredient of Theorem~\ref{thm:main2} is that it holds at an
infinitesimal level, i.e., for the linearized map: a linear map should act on
Lebesgue measure by multiplying it according to the absolute value of its
determinant.

Here is the formalized statement:
\begin{lstlisting}
lemma add_haar_image_linear_map
[normed_group E] [normed_space ℝ E] [measurable_space E] [borel_space E]
[finite_dimensional ℝ E] (μ : measure E) [is_add_haar_measure μ]
(f : E →ₗ[ℝ] E) (s : set E) :
μ (f '' s) = ennreal.of_real (abs f.det) * μ s
\end{lstlisting}

The proof of this theorem splits into two steps.

Let us first prove it on $E = \R^n$, with its standard product measure. A
linear map $f$ on this space corresponds in a canonical way to a matrix $M$.
Gaussian elimination shows that $M$ can be written as the product of
transvection matrices (i.e., matrices with ones on the diagonal and at most
one non-zero off-diagonal coefficient) and a diagonal matrix. As the
statement to be proved is clearly stable under multiplication (as the
determinant of a product is the product of the determinants), it therefore
suffices to check it for transvections and for diagonal matrices. For
diagonal matrices, it follows readily from the fact that the measure is a
product measure and from the elementary $1$-dimensional situation. For
transvections (whose determinant is $1$), we should show that a transvection
preserves Lebesgue measure. This follows from a straightforward computation
using Fubini to single out the coordinate at which there is a nonzero entry
in the matrix.

The key argument of this step is thus Gaussian elimination, which was not yet
proved  in \mathlib before this project and was formalized with this goal in
mind. This is probably the most unexpected outcome of this project!

Let us now turn to the case of a general finite-dimensional vector space $E$,
with a general Lebesgue measure $\mu$. There is no canonical basis, and
therefore no way to identify a linear map with a matrix, and moreover a
general Lebesgue measure has no product structure a priori, so that the above
argument does not make sense. However, one can choose a basis, which yields
an isomorphism $A$ between $\R^n$ and $E$ for some $n$. Let $g \coloneqq
A^{-1} \circ f \circ A$ be the conjugate of $f$ under this isomorphism. Its
determinant coincides with that of $f$, and moreover the first step applies
to $g$. To conclude, it suffices to show that the image under $A$ of the
standard Lebesgue measure on $\R^n$ coincides with $\mu$ (or a scalar
multiple of $\mu$). We deduce this from uniqueness of Lebesgue measure up to
scalar multiplication, which was already available in \mathlib in the right
generality: two Haar measures on a locally compact group are multiples of
each other. This is a nontrivial fact, which had fortunately already been
formalized by van Doorn with a totally unrelated application in mind,
see~\cite{van_doorn_haar}.

\begin{rmk}
Once the above theorem is available, one deduces that the volume of balls
behaves like $\mu(B(x,r)) = r^d \mu(B(0,1))$ where $d$ is the dimension of
the space. This fact makes it possible to apply the Vitali covering theorem
(Theorem~\ref{thm:vitali} below) to $\mu$, and is therefore fundamental. Note
that it is obvious for the standard product measure in the Euclidean space
$\R^n$, but not so obvious in a general normed vector space with a general
Lebesgue measure.
\end{rmk}

\section{Covering theorems in measure theory}
\label{sec:covering}

There is a huge variety of covering theorems for measure theory in the
literature. Among them, the two most prominent ones are probably the Vitali
and Besicovitch covering theorems that we will now describe and that we have
formalized for the current project.

Here is a version of the Vitali covering theorem:
\begin{thm}
\label{thm:vitali} In a metric space $X$, consider a locally finite measure $\mu$
which is doubling: there exists $C>0$ such that, for any $x,r$, then
$\mu(B(x, 2r)) \le C \mu(B(x,r))$. Consider a set of balls $t$ with uniformly
bounded radii, and a set $s\subseteq X$ at which the family $t$ is fine, i.e., every
point of $s$ belongs to balls in $t$ with arbitrarily small radius. Then
there exists a disjoint subfamily of $t$ covering almost all $s$.
\end{thm}

Here is a version of the Besicovitch covering theorem:
\begin{thm}
Let $E$ be a finite-dimensional real normed vector space, and $s$ a subset of
$E$. Consider a set $t$ of balls such that, for any $x\in s$, there exist
arbitrarily small radii $r$ such that $B(x,r) \in t$. Let $\mu$ be a sigma-finite
measure. Then there exists a disjoint subfamily of $t$ covering almost all
$s$.
\end{thm}

The assumptions of the Besicovitch theorem on the measure are weaker than
those of the Vitali theorem, as they do not assume any doubling property. On
the other hand, the former theorem requires stronger geometric assumptions
(in this formulation, a finite-dimensional real normed vector space instead
of a general metric space). The Besicovitch theorem is harder to prove and
often more powerful than the Vitali theorem. The book~\cite{federer} is a
standard reference for these theorems and several powerful extensions, which
was used for the formalization.

Both theorems are consequences of purely combinatorial results, from which
the measure theoretic versions are then deduced. For instance, the
deterministic result leading to the Besicovitch covering theorem is the
following statement:
\begin{thm}
Let $E$ be a finite-dimensional real normed vector space. There exists a
constant $N=N(E)$ with the following property. Consider a set $s$, and a set
of balls $t$ with uniformly bounded radii such that any point in $s$ is the
center of some ball in $t$. Then one may find disjoint subfamilies
$t_1,\dotsc, t_N$ of $t$ which still cover $s$.
\end{thm}
To deduce the measure-theoretic version from this deterministic version, one
picks one of the subfamilies covering a proportion $1/N$ of $s$, say $t_i$,
and then works inductively on the subset of $s$ that it still to be covered.

The best value of the constant $N$ has been determined in~\cite{furedi_loeb}:
it is the maximal number of points one can put inside the unit ball of radius
$2$ under the condition that their distances are bounded below by $1$. This
is the version we have formalized in \mathlib.

Let us mention that this theorem (and the combinatorial theorem implying the Vitali
covering theorem) are proved using transfinite induction (i.e., an induction indexed by
ordinals). Fortunately, ordinals were already available in \mathlib before the start of
this project.

\medskip

The book~\cite{federer} introduces a formalism to deduce consequences of
covering theorems in a uniform way, independently of the covering theorem.
This formalism is often deemed too abstract by mathematicians, but
it turns out to be extremely well suited to formalization. Let us say that a
family of sets $(\mathcal{F}_x)_{x\in X}$ in a metric measured space $X$ is a
Vitali family if it satisfies the following property: consider a (possibly
non-measurable) set $s$, and for any $x$ in $s$ a subfamily $F_x$ of
$\mathcal{F}_x$ containing sets of arbitrarily small diameter. Then one can
extract from $\bigcup_{x \in s} F_x$ a disjoint subfamily covering almost all
$s$.

In this language, the Vitali covering theorem says that one gets a Vitali
family by taking for $\mathcal{F}_x$ the balls that contain $x$, in a space
where the measure $\mu$ is doubling. And the Besicovitch covering theorem
states that, in a finite-dimensional real vector space, one gets a Vitali
family by taking for $\mathcal{F}_x$ the balls centered at $x$.

The fundamental theorem on differentiation of measures is the following. On a
metric space with a measure $\mu$, consider a Vitali family $\mathcal{F}$ and
another measure $\rho$. Then, for $\mu$-almost every $x$, the ratio $\rho(a)/
\mu(a)$ converges when $a$ shrinks to $x$ along the Vitali family
$\mathcal{F}_x$, towards the Radon-Nikodym derivative of $\rho$ with respect
to $\mu$. We have formalized this theorem, as follows:
\begin{lstlisting}
theorem vitali_family.ae_tendsto_rn_deriv
[sigma_compact_space α] [borel_space α]
{μ : measure α} [is_locally_finite_measure μ] (v : vitali_family μ)
(ρ : measure α) [is_locally_finite_measure ρ] :
∀ᵐ x ∂μ, tendsto (λ a, ρ a / μ a) (v.filter_at x) (nhds (ρ.rn_deriv μ x))
\end{lstlisting}

Note that the convergence when a set in a Vitali family shrinks to a point
(i.e., its diameter tends to $0$) is not one of the standard convergences in
mathematics (and Federer discusses it at length and introduces a special
notation for it) but it fits perfectly well within the framework of \mathlib
where all notions of convergence are expressed with the single notion of
filter, as advocated by~\cite{analysis_HOL}.

Once this theorem is formalized, versions in the context of the Vitali and
the Besicovitch covering theorems readily follow. In these respective
contexts, they make it possible to prove that almost every point $x$ of a set
$s$ is a Lebesgue density point of $s$, i.e., $\mu(s \cap B(x,r))/\mu(s) \to
1$ as $r \to 0$. This fact plays a key role in the proof of
Theorem~\ref{thm:main2} to prove that $x\mapsto f'(x)$ is almost everywhere
measurable (indeed, it is measurable when restricted to the set of Lebesgue
density points).

\section{Polish spaces and descriptive set theory}

A Polish space is a topological space which is second-countable and on which
there exists a complete metric space structure inducing the given topology. A
good reference on Polish spaces and descriptive set theory is~\cite{kechris}.
This definition may seem strange at first: one could instead require to have
a complete second-countable metric space. However, in many applications,
there is no natural distance, and the only relevant information is the
topology. For instance, the extended reals $[-\infty, \infty]$ are a Polish
space, but they don't have a canonical metric to work with.

This slightly awkward definition makes Polish spaces slightly awkward to use
in proof assistants, as it refers to the mere existence of a nice metric but
without providing it. In the completely classical framework of \mathlib, this
is not a real issue as one can use choice to pick such an arbitrary nice
metric. The definition is formalized as follows:
\begin{lstlisting}
class polish_space (α : Type*) [h : topological_space α] : Prop :=
(second_countable [] : second_countable_topology α)
(complete : ∃ m : metric_space α,
    m.to_uniform_space.to_topological_space = h ∧
    @complete_space α m.to_uniform_space)
\end{lstlisting}

To construct a nice metric space structure on a Polish space, one uses the
following incantation in proofs: \verb+letI := upgrade_polish_space α+. It
endows the Polish space $\alpha$ with a metric space structure which is
registered in the typeclass system as complete and second-countable, and
moreover the topology associated to this metric is definitionally equal to
the given topology. This makes it possible to work smoothly with Polish
spaces as one would do in a non-formalized setting.

\medskip

An important theme in descriptive set theory is to start with a Polish space
and modify its topology to get better behavior while retaining Polishness.
For instance, if a map between two Polish spaces is measurable, then one can
construct a finer Polish topology on the source space for which the map
becomes continuous. This makes it possible to deduce results for measurable
maps from results for continuous maps. Unfortunately, this kind of argument
does not interact well with the typeclass system, in which each type is
supposed to be endowed with at most one typeclass of each kind, and in
particular at most one topology. Fortunately, there is a way to override
typeclass inference and provide explicitly the topology one would like to use
(at the cost of added verbosity and reduced readability), by prefixing a command with
\verb+@+ and then giving explicitly all its arguments, including the implicit
and typeclass arguments. For instance, let
us give the formalized version of the following basic (but nontrivial)
statement: given a Polish space and countably many finer Polish topologies,
there exists another Polish topology which is finer than all of them.
\begin{lstlisting}
lemma exists_polish_space_forall_le {ι : Type*} [encodable ι]
[t : topological_space α] [polish_space α]
(m : ι → topological_space α) (hm : ∀ n, m n ≤ t)
(h'm : ∀ n, @polish_space α (m n)) : ∃ (t' : topological_space α),
  (∀ n, t' ≤ m n) ∧ (t' ≤ t) ∧ @polish_space α t'
\end{lstlisting}
This fact is proved as follows. First, one checks (by constructing a suitable
complete metric) that a countable product of Polish spaces is Polish.
Consider then the infinite product
$Y=\alpha^{\N}$, where the $n$-th copy of $\alpha$ is endowed with the $n$-th
topology $m_n$. Then $Y$ is Polish, therefore the pullback of its topology
under the diagonal embedding is also Polish and it satisfies all the required
properties.

Another trick that proves useful in this kind of argument is to use a type
synonym, i.e., a copy of a type with a different name. As typeclass inference
does not unfold definitions, the type synonym is not endowed by default with any
topology or metric, and one can register new instances that will not conflict
with the original ones. As an example, let us sketch the proof that an open
subset of a Polish space is Polish:
\begin{lstlisting}
lemma is_open.polish_space [topological_space α] [polish_space α]
{s : set α} (hs : is_open s) : polish_space s
\end{lstlisting}
One endows $\alpha$ with a distance for which it is complete and
second-countable. Then the open subset $s$ of $\alpha$ has the induced
topology, which is second-countable, and it also inherits the restricted
distance. But in general it is not complete for the restricted distance
(think of the interval $(0, 1) \subset \R$). One should therefore use another
distance. A suitable formula for it is
\begin{equation*}
  d'(x,y) = d(x,y) + \abs*{1/d(x, \alpha \setminus s) - 1/d(y, \alpha \setminus s)}.
\end{equation*}
As it blows up close to the boundary of $s$, one can check that this is a
complete distance on $s$, defining the same topology as the original topology
and therefore proving that it is Polish. For formalization purposes, we do
not put this distance on $s$ (which would then have two competing metric
space structures and would force us to use the \verb+@+ version of statements everywhere),
but instead on a type synonym \verb+complete_copy s+. Then
there is no difficulty to check that \verb+complete_copy s+ is Polish. As
this new distance defines the same topology as the original one, the identity
from \verb+s+ to \verb+complete_copy s+ is a homeomorphism, hence the fact
that \verb+s+ itself is Polish follows.

\medskip

The Lusin-Souslin theorem states that the image of a measurable set under a
measurable injective map on a Polish space is still measurable. Here is the
formalized statement:
\begin{lstlisting}
theorem measurable_set.image_of_measurable_inj_on
[topological_space γ] [polish_space γ] [measurable_space γ]
[borel_space γ] [topological_space β] [t2_space β] [measurable_space β]
[borel_space β] [second_countable_topology β] {s : set γ} {f : γ → β}
(hs : measurable_set s) (f_meas : measurable f) (f_inj : set.inj_on f s) :
measurable_set (f '' s)
\end{lstlisting}

Its proof builds on the techniques we have sketched above, but it is
considerably more involved. We will not get into more details.
Let us just mention that it is first proved assuming that $f$ is continuous,
and then generalized to a measurable $f$ by using the trick to modify the
topology to turn a measurable map into a continuous map. A key notion in the
proof is that of an analytic set, i.e., the image of a Polish space under a
continuous map, and a key result is that if two analytic sets are disjoint,
then they are contained in disjoint measurable sets (this result is called
the Lusin separation theorem). As far as the author knows, there is no known
proof of the Lusin-Souslin theorem which does not go through a study of
analytic sets, even though these sets do not appear in the conclusion of the
theorem.

\section{Conclusion}

We have described the formalization of
the change of variables formula in integrals, in an advanced version. One
interest of this theorem from the point of view of formalization is that it
involves several areas of mathematics that are often considered quite
independent, but that need to interact seamlessly here -- as is often the
case in advanced mathematics that mix basic results from several areas.

We have explained how the development model of \mathlib has made this project
reasonable, as well as its general philosophy: things should be done right,
in the greatest level of generality, and without taking shortcuts. A lot of
refactors are needed to reach this goal. This is possible in \mathlib since
it is a mono-repository project, without a lot of outside users, and would be
harder for more mature projects. We may hope that basic definitions stabilize
with time, but we are clearly not there yet: the definition of a group was
changed less than one year ago to cope with a definitional equality issue in
tensor products of abelian groups seen as $\Z$-modules, that showed up when
doing advanced mathematics in the liquid tensor experiment. This kind of
agile development is clearly a strength of \mathlib currently, but once it
reaches a critical size other strategies will need to be devised to make sure
it can be used in a more stable way by other projects.

\medskip

The change of variables formula is one of the tools to implement de Rham
cohomology. While this project was being done, other necessary tools have
been formalized independently, for other projects: homological algebra was
developed for the liquid tensor experiment, and vector bundles on manifolds
were developed for the sphere eversion project. This means that de Rham
cohomology is now a reasonable target!

\bibliography{biblio}
\bibliographystyle{splncs04}

\end{document}